# Cluster Correlations in the Zel'dovich Approximation

Stefano Borgani[1,2], Peter Coles[3] and Lauro Moscardini[4]
[1] *INFN Sezione di Perugia, Dipartimento di Fisica dell'Università, via A. Pascoli, I–06100 Perugia, Italy*
[2] *Scuola Internazionale Superiore di Studi Avanzati, SISSA, via Beirut 2-4, I–34013 Trieste, Italy*
[3] *Astronomy Unit, School of Mathematical Sciences, Queen Mary and Westfield College, Mile End Road, London E1 4NS*
[4] *Dipartimento di Astronomia, Università di Padova, vicolo dell'Osservatorio 5, I–35122 Padova, Italy*

2 December 1993

**ABSTRACT**
We show how to simulate the clustering of rich clusters of galaxies using a technique based on the Zel'dovich approximation. This method well reproduces the spatial distribution of clusters obtainable from full $N$–body simulations at a fraction of the computational cost. We use an ensemble of large–scale simulations to assess the level and statistical significance of cluster clustering in open, tilted and flat versions of the Cold Dark Matter (CDM) model, as well as a model comprising a mixture of Cold and Hot Dark Matter (CHDM). We find the open and flat CDM models are excluded by the data. The tilted CDM model with a slight tilt is in marginal agreement, while larger tilt produces the right amount of clustering; CHDM is the best of all our models at reproducing the observations of cluster clustering. We find that *all* our models display a systematically weaker relationship between clustering length and mean cluster separation than seems to be implied by observations. We also note that the bias factor, defined by the ratio of cluster correlations to the linear mass correlations is not constant in any of the models, showing that one needs to be very careful when relating cluster clustering statistics to primordial density fluctuations.

## 1 INTRODUCTION

Redshift surveys of catalogues of rich clusters of galaxies provide ample evidence that galaxy clusters are much more strongly correlated than galaxies themselves (Bahcall & Soneira 1983; Klypin & Kopylov 1983; Postman, Huchra & Geller 1992; Dalton et al. 1992; Nichol et al. 1992; Peacock & West 1992; Scaramella, Vettolani & Zamorani 1993; Nichol, Briel & Henry 1993). The strong clustering of clusters makes them potentially powerful probes of the large–scale structure of the Universe, because their clustering can be detected with high statistical significance on scales where the galaxy-galaxy correlations are so small as to be undetectable.

There are some problems, however, in calculating the amplitude of cluster clustering expected in any specific cosmological scenario. The usual technique for galaxy clustering involves $N$–body numerical simulations which are computationally expensive to perform on the scale needed to simulate cluster formation (White et al. 1987; Bahcall & Cen 1992; Klypin & Rhee 1993, hereafter KR93; Croft & Efstathiou 1993). Moreover, in order to assess correctly the statistical significance of the results obtained one should ideally perform a large ensemble of simulations and look at the variation of statistical measures obtained across the ensemble. Alternative analytical approaches based upon the properties of peaks in the linear density field (Kaiser 1984; Bardeen et al. 1986; Bardeen, Bond & Efstathiou 1987; Coles 1989; Lumsden, Heavens & Peacock 1989; Borgani 1990; Holtzman & Primack 1993) are of uncertain reliability because of the simplifying assumptions they make about the cluster formation process.

In this paper, we propose an alternative method for simulating cluster samples in cosmological scenarios which is based on the Zel'dovich approximation (hereafter ZA; Zel'dovich 1970). The virtue of this method is that the Zel'dovich approximation, if used appropriately, reproduces the spatial distribution of matter found in fully dynamical $N$–body simulations for a wide variety of models at a fraction of the computational cost. We are therefore able to present a study of a large ensemble of realizations of cluster distributions from various cosmological scenarios with realistic error statistics.

Mann, Heavens & Peacock (1993) have discussed the application of ZA to the cluster distribution in some detail. Although their analytical approach differs from ours in many respects, they do conclude that ZA gives a fair description of gravitational clustering on scales probed by clusters, i.e. $\gtrsim 10\,h^{-1}$ Mpc. (As usual $h$ is the Hubble parameter in units of 100 km s$^{-1}$ Mpc$^{-1}$.)

We outline our method in the next section and then present an analysis of the results in Section 3. Section 4 contains a summary of our conclusions.

## 2  THE ZEL'DOVICH TECHNIQUE

Our method for simulating the formation of clusters is based on the Zel'dovich approximation (Zel'dovich 1970; Shandarin & Zel'dovich 1989). Let $\mathbf{x}(t)$ be the comoving position of a particle, which is related to its proper coordinate $\mathbf{r}(t)$ by $\mathbf{x}(t) = \mathbf{r}(t)/a(t)$, where $a(t)$ is the usual cosmological expansion factor. The general solution to the equations of motion for the fluid particles can be written as a mapping of Eulerian-to-Lagrangian coordinates:

$$\mathbf{x}(\mathbf{q},t) = \mathbf{q} + \mathbf{s}(\mathbf{q},t), \tag{1}$$

where $\mathbf{q}$ and $\mathbf{x}$ are the initial (Lagrangian) and the final (Eulerian) particle positions respectively, and $\mathbf{s}(\mathbf{q},t)$ is the displacement vector. The Zel'dovich approximation amounts to factorizing the $t$- and $\mathbf{q}$-dependence in the expression of $\mathbf{s}(\mathbf{q},t)$. For a collisionless, self-gravitating fluid in an expanding background it turns out that

$$\mathbf{s}(\mathbf{q},t) = b(t)\,\nabla\psi(\mathbf{q}), \tag{2}$$

where $b(t)$ is the growing mode for the linear evolution of density fluctuations and $\psi(\mathbf{q})$ describes the gravitational potential. The gravitational potential is related to the density fluctuation field by Poisson's equation

$$\nabla^2\psi(\mathbf{q}) = \frac{1}{a(t)}\delta(\mathbf{q}). \tag{3}$$

According to eqs.(1) and (2), the Zel'dovich approximation describes straight-line trajectories, on which particles move with constant velocity,

$$\mathbf{v} = \dot{x} = \dot{b}\,\nabla\psi(\mathbf{q}). \tag{4}$$

Thus, particles do not feel any gravitational field and, although they are correctly moved by gravity at the beginning, they are not held within the minima of the potential field at later times. Structures form, but are later erased when particle trajectories cross. Despite its simplicity and its limited treatment of dynamics inside high-density regions, the ZA is known to represent an efficient tool to describe gravitational clustering in the mildly non-linear regime. Coles, Melott & Shandarin (1993) compared ZA clustering with results based on "exact" PM $N$-body simulations. They found that, after suitable filtering of the small wavelength modes from the initial linear fluctuation spectrum, shell-crossing is reduced to a level at which the ZA provides a surprisingly good description of clustering. The required truncation of the initial spectrum is to erase the fluctuation modes with wavelength $\lambda \lesssim R$, where $R$ is to be chosen so that the variance at this scale does not exceed unity, $\sigma_R^2 \lesssim 1$. It turns out therefore, that the performance of the ZA should be about optimal on cluster scales ($\gtrsim 10\,h^{-1}$ Mpc), where the *rms* density contrast is of order unity and clustering is consequently in the quasi-linear regime.

Before assigning the initial (linear) density field, we convolve the power-spectrum with a Gaussian window function, $W_R(k) = \exp(-R^2k^2/2)$. This choice of the window function has been shown by Melott, Pellman & Shandarin (1993) to optimize the ability of ZA to follow non-linear clustering. The value of the smoothing scale $R$ is chosen so that the mass contained inside the window volume is of the same order of that of a cluster. Since $V_R = (2\pi R^2)^{3/2}$ is the volume encompassed by the Gaussian window and taking $M_{cl} \simeq 10^{15} M_\odot$ for the typical cluster mass-scale, we choose $R = 5\Omega_0^{-1/3}\,h^{-1}$ Mpc for the smoothing scale.

Once we have assigned the smoothed power spectrum, we solve for the potential $\psi(\mathbf{q})$ on $128^3$ grid points for a $L = 640\,h^{-1}$ Mpc simulation box. This very large box size allows us to sample up to the largest scales involved in the available cluster redshift surveys. The clustering developed by ZA is followed by putting a particle on each grid point and shifting it according to eq.(2). After particles have been moved, we reconstruct the density field on the grid through a TSC interpolating scheme (e.g. Hockney & Eastwood 1981). We then identify clusters with peaks of the *evolved* density field according to the following prescription. For a chosen value of the mean cluster separation, $d_{cl}$, the expected number of clusters is $N_{cl} = (L/d_{cl})^3$. Then, if $M_i$ is the mass of the $i$-th density peak, the number of clusters to be associated with it is $N_i = M_i/\bar{M}$, where the *critical* mass $\bar{M}$ is chosen so to have $\sum_i N_i = N_{cl}$. This prescription amounts to a sort of biasing, since peaks with mass below $\bar{M}$ do not generate clusters, while very high peaks contains more than one cluster. The values of $\bar{M}$ for each model are displayed in Table 1, once we take $d_{cl} = 40\,h^{-1}$ Mpc, which corresponds to the mean separation for $R \geq 0$ Abell clusters. These values can be compared with the mass threshold ($M > 1.8 \times 10^{14}\,h^{-1} M_\odot$) estimated by Bahcall & Cen (1993) in their study of the

**Table 1.** The models. Columns 2 to 5 contain the total density parameter $\Omega_0$ and the contributions to it due to CDM, HDM and baryons, respectively. Column 6 contains the post–inflationary spectral index $n$, column 7 is the Hubble parameter $h$, column 8 is for the linear biasing parameter $b$, column 9 is for the variance $\sigma_R^2$ inside the Gaussian window of $5\Omega_0^{-1/3}\,h^{-1}\,{\rm Mpc}$ radius, and column 10 is for the cluster mass threshold $\bar{M}$ (see text).

| Model | $\Omega_0$ | $\Omega_{CDM}$ | $\Omega_{HDM}$ | $\Omega_{bar}$ | $n$ | $h$ | $b$ | $\sigma_R^2$ | $\frac{\bar{M}}{10^{14} M_\odot}$ |
|---|---|---|---|---|---|---|---|---|---|
| SCDM | 1.00 | 0.90 | 0.00 | 0.10 | 1.0 | 0.5 | 1.0 | 0.59 | 1.2 |
| OCDM | 0.20 | 0.19 | 0.00 | 0.01 | 1.0 | 1.0 | 1.0 | 0.10 | 0.2 |
| TCDM | 1.00 | 0.90 | 0.00 | 0.10 | 0.8 | 0.5 | 1.5 | 0.44 | 1.1 |
| CHDM | 1.00 | 0.60 | 0.30 | 0.10 | 1.0 | 0.5 | 1.5 | 0.31 | 1.0 |

Abell cluster mass function. Since ZA has troubles to keep particles confined within potential wells, it is not astonishing that we generally underestimate cluster masses.

Our cluster identification procedure is one way to parametrize our ignorance about what is really happening inside individual peaks. Since the smoothing procedure erases any dynamical effects below the scale $R$, it is appropriate to add a random displacement to the cluster positions on the grid with *rms* value equal to $R$. This eliminates any discrete lattice effects on the evaluation of the correlation function at small scales ($\lesssim 10\,h^{-1}\,{\rm Mpc}$) and leaves the large–scale behaviour (in which we are most interested) substantially unchanged. For each model, we take an ensemble of 20 different random realizations, so that we are reasonably sure to deal properly with the effect of 'cosmic variance', which could be particularly severe for clustering models with substantial large–scale power.

In order to test the reliability of the cluster selection procedure, and to get a feel for the effect of this procedure upon clustering statistics, we also attempted to associate only one cluster per peak, so to select only the $N_{cl}$ highest peaks. With this method, we are not able to reproduce the $N$–body results on small ($\lesssim 20\,h^{-1}\,{\rm Mpc}$) scales (see below), but the results do not change appreciably on large scales.

The ability of ZA to describe the large–scale statistics of the cluster distribution has been already discussed by several authors. Batuski, Melott & Burns (1987), Postman et al. (1989) and Plionis, Valdarnini & Jing (1992) used ZA to generate clustered point distributions to be compared with the observed cluster samples. The main difference between these previous approaches and ours is that the above authors were trying to find a linear power–spectrum which produces a cluster distribution with nearly the same statistics as observed. They use ZA just to move those particles, associated to high density peaks of the linear field, to be associated with clusters. More recently, Mann et al. (1993) used an analytical approach to study the cluster distribution implied by ZA. Again, they identify clusters as peaks of the linear density field, then use ZA to move their Lagrangian positions. The problem with this is that we find very little correspondence between peaks of the initial density field and those in the field evolved by ZA; see also Coles et al. (1993). Furthermore, although a fully analytical approach should be in general preferred to numerical simulations, it does not allow an easy investigation of higher–order correlation properties and of the effect of observational biases (sample geometries, selection effects, etc.).

We use ZA to generate clusters simulations from physical power spectra to check their ability to provide a reliable large–scale clustering description. Furthermore, we identify clusters from peaks of the evolved field, which are expected to correspond more closely to the sites where non–linear gravity generates galaxy clusters.

## 3 ANALYSIS OF THE RESULTS

We have generated ZA simulations for four different cosmological models: (1) Standard CDM model (SCDM), with amplitude normalised to the COBE data (Smoot et al. 1992), i.e. $b = 1$ for the linear biasing parameter; (2) Open CDM (OCDM), with $\Omega_0 = 0.2$; (3) Tilted CDM (TCDM), with spectral index $n = 0.8$ for the post–inflationary spectrum and $b = 1.5$; (4) A 'mixed' model (CHDM), with about 30% HDM. A detailed description of the relevant parameters of each of our models is given in Table 1; transfer functions were taken from Holtzmann (1989). [For further discussion of tilted CDM, see Cen et al. (1992), Adams et al. (1993) and Tormen et al. (1993); for CHDM see Klypin et al. (1993) and reference therein.] Our OCDM model does not have a cosmological constant to make the space–time flat; this, however, is thought to have little effect on the final distribution (see also Bahcall & Cen 1992). The variances $\sigma_R^2$ at the smoothing scale $R$ for each of the models are also reported in Table 1. We note that $\sigma_R^2 < 1$, so that we are confident that we are following the gravitational clustering in a regime where ZA will work.

In Figure 1 we show the projected distribution of clusters with $d_{cl} = 40\,h^{-1}\,{\rm Mpc}$ in slices of thickness a $100\,h^{-1}\,{\rm Mpc}$ from one realization of each model. The phase assignment is the same for all the plots, so that a fair comparison of the structures in the different models is possible. Visually, there are clear differences between the models: the SCDM model produces a rather smooth cluster distribution; TCDM and CHDM have more obvious structure; OCDM produces the largest structures of all, with voids and superclusters on scales even greater than $100\,h^{-1}\,\mathrm{Mpc}$.

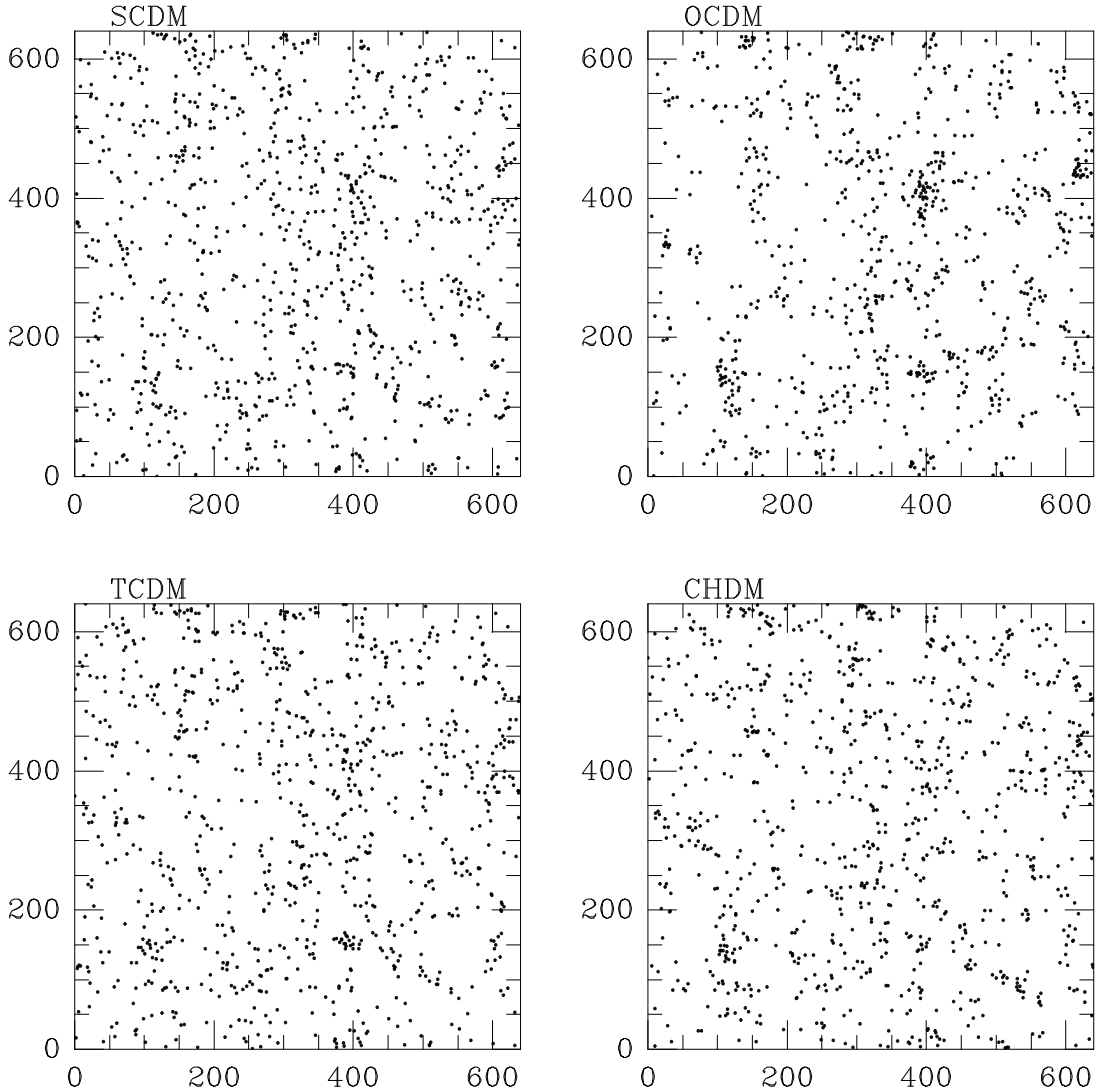

**Figure 1.** The cluster distribution projected from a $100\,h^{-1}\,\mathrm{Mpc}$ slice for one realization of each model. The box size is $640\,h^{-1}\,\mathrm{Mpc}$. The plotted clusters correspond to a mean separation $d_{cl} = 40\,h^{-1}\,\mathrm{Mpc}$. The phase assignment is the same for all the models.

We evaluated the two–point correlation function for the cluster simulations by adopting the direct estimator

$$\xi_{cc}(r) = \frac{N_{cc}}{n_c^2 L^3 \delta V} - 1, \qquad (5)$$

where $N_{cc}$ is the number of cluster–cluster pairs with separation between $r - \Delta/2$ and $r + \Delta R/2$, $n_c$ is the average cluster number density and $\delta V$ is the volume of the spherical shell between the two limiting bin radii. All clusters of each simulation are used and periodic boundary conditions are imposed. The results for each model are obtained by averaging over the 20 random realizations and errors in $\xi_{cc}(r)$ are determined from the *rms* scatter across this ensemble. Our errors therefore contain an estimate of the cosmic variance, which the usual bootstrap resampling errors (Ling, Frenk & Barrow 1986) do not.

In Figure 2 we compare our results for CHDM clusters with $d_{cl} = 40\,h^{-1}\,\mathrm{Mpc}$ with the output of $N$–body simulations of $R \geq 0$ Abell clusters by KR93. Notice that, unsurprisingly, the two results disagree at small separations, corresponding to the smoothing scale $R$ of the linear field, but are within $1\sigma$ errors at large scales. Notice that the errors on the KR93 results are Poisson estimates, while ours represent the *rms* over an ensemble of 20 simulations. In doing this comparison, we should bear in mind that the KR93 results are obtained by averaging within two realizations in a $200\,h^{-1}\,\mathrm{Mpc}$ box. The much larger size of each of our 20 realizations ($640\,h^{-1}\,\mathrm{Mpc}$) allows a much better sampling of long wavelength modes, especially for the CHDM spectrum, which has a substantial amount of large–scale power. This is also the reason for the positive correlation shown by $N$-body results at $\sim 50\,h^{-1}\,\text{Mpc}$, not reproduced by our simulations (see Figure 4 below). In any case, the agreement is encouraging and suggests that our method does indeed well represent the process of cluster formation.

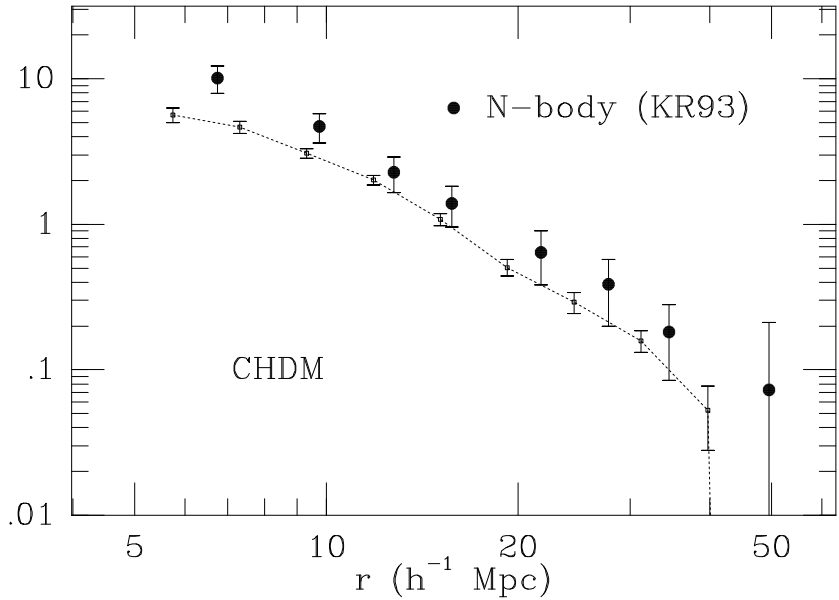

**Figure 2.** Comparison between our CHDM results on $\xi(r)$ and that obtained in $N$-body simulations by KR93 for $R \geq 0$ Abell clusters ($d_{cl} = 40\,h^{-1}\,\text{Mpc}$).

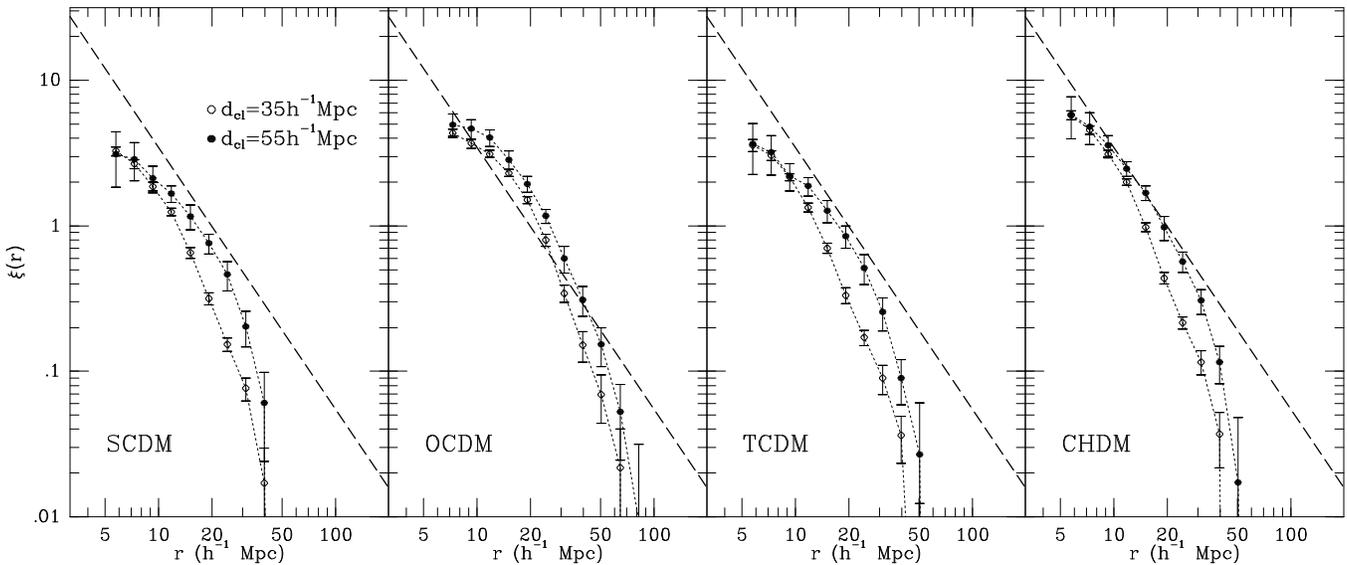

**Figure 3.** The two-point cluster correlation function for our models. In each panel, open symbols correspond to clusters with mean separation $d_{cl} = 35\,h^{-1}\,\text{Mpc}$, i.e. that of APM clusters (Dalton et al. 1992), while filled dots are for $d_{cl} = 55\,h^{-1}\,\text{Mpc}$, i.e. $R \geq 1$ Abell clusters (Postman, Huchra & Geller 1992). The dashed straight line is the power-law $\xi(r) = (20\,h^{-1}\,\text{Mpc}/r)^{1.8}$, which is claimed to provide a good fit to the rich Abell cluster data.

In Figure 3 we plot $\xi(r)$ for our four models. Some interesting points are immediately apparent from this figure. First of all, notice than none of these models actually produces a pure power-law shape for $\xi_{cc}(r)$ over the whole range of length scales: they all exhibit considerable curvature. There is some tentative evidence for this kind of behaviour also in the observational data (Scaramella et al. 1993). Increasing the mean separation of clusters means selecting rarer, higher peaks, with a corresponding increase of the correlation amplitude. The SCDM model produces a $\xi(r)$ which is too low and which has a sharp break at about $40\,h^{-1}\,\text{Mpc}$. The OCDM model produces clustering that is too strong, with a correlation length for the APM-like clusters of about $20\,h^{-1}\,\text{Mpc}$, much higher than the $r_0 \simeq 13\,h^{-1}\,\text{Mpc}$ claimed by Dalton et al. (1992). The TCDM model with $n = 0.8$ shows only a marginal agreement with the observations. In order to check the behaviour of models with larger tilt, we also ran simulations with $n = 0.7$. The resulting correlation function (not shown here) is higher and in better agreement; note however that these models are strongly constrained by a joint analysis of the large–scale velocity fields and the COBE results (Moscardini et al. 1993). On the other hand, CHDM model produces roughly the right level of clustering in both cluster populations over the whole range of scales considered.

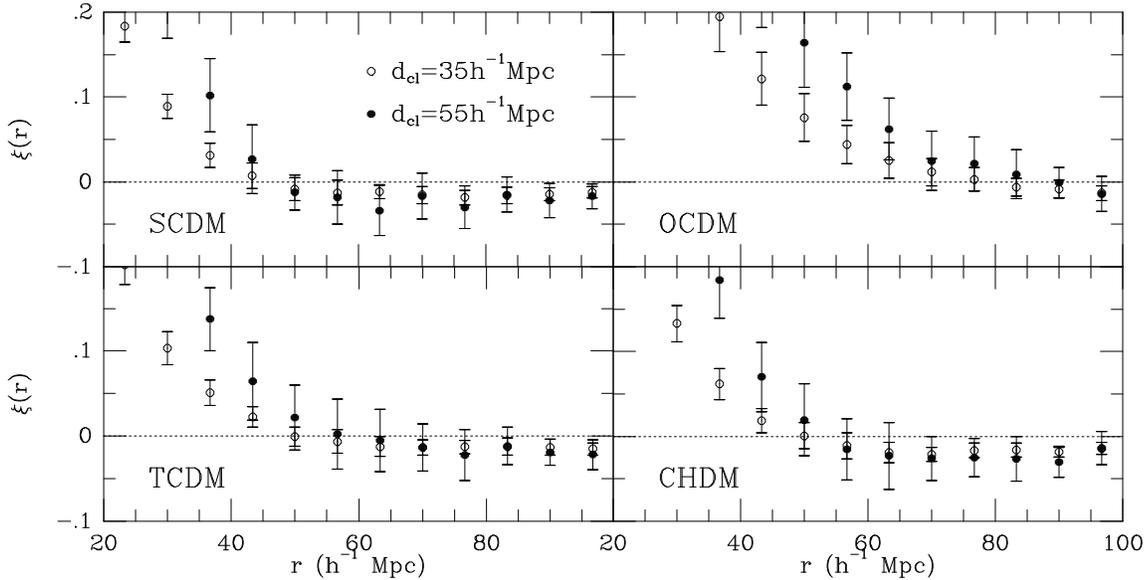

**Figure 4.** The first–zero crossing of $\xi(r)$. Open and filled dots are for $d_{cl} = 35\,h^{-1}$ Mpc and $d_{cl} = 55\,h^{-1}$ Mpc clusters, respectively.

Our results can be compared with those based on $N$–body simulations already presented in the literature [see Fig.2 in Croft & Efstathiou (1993) for the SCDM and OCDM model; no similar comparisons are available for TCDM]. In all the cases for which a close comparison is possible, the ZA results are in good agreement with those from $N$–body experiments.

KR93 have suggested that $r_c$, the first zero–crossing of $\xi_{cc}(r)$, might be a useful statistic with which to discriminate between models. Several studies of cluster clustering suggests that $r_c = (50 \pm 10)\,h^{-1}$ Mpc (Postman et al. 1992; Cappi & Maurogordato 1992; Peacock & West 1992; Dalton et al. 1992; Scaramella et al. 1993). The behaviour of our simulated $\xi_{cc}(r)$ around the zero–crossing are shown in Figure 4. The lack of power of SCDM means that $\xi_{cc}(r)$ is consistent with zero at $r \gtrsim 40\,h^{-1}$ Mpc. A shift toward larger $r_c$ values occurs for the TCDM and CHDM models, which have $r_c \simeq 50\,h^{-1}$ Mpc, in agreement with observations. The OCDM model gives $r_c \simeq 80\,h^{-1}$ Mpc, which seems to be significantly larger than the value suggested by observational data. From these plots, it is also apparent that $r_c$ does not vary significantly with $d_{cl}$, as the correlation length does, but only on the shape of the linear power–spectrum. Because it does not depend upon the richness criterion used to select the cluster samples, $r_c$ might, in principle, be a useful test statistic. On the other hand, the task of estimating $r_c$ is complicated by uncertainties in the mean number–density of objects. Moreover, the ensemble error in $r_c$ is not negligible and this may be exacerbated when realistic sample geometries and selection functions are employed. This result may also be influenced by the way numerical simulations sample long wavelength modes. To check this out, we ran simulations with half the box size. In such simulations, we get a smaller $r_c$, but the size of the shift depends upon the amount of large–scale power. For example, in the OCDM model we get $r_c \simeq 60\,h^{-1}$ Mpc for a $320\,h^{-1}$ Mpc box. The combination of numerical and sampling difficulties makes it doubtful whether this statistic is of any practical usefulness.

Another possible test involving rich clusters is the richness–clustering relationship discussed by Bahcall & Burgett (1986) and Bahcall (1988). This relationship, recently questioned by Scaramella et al. (1993), can be cast into a dependence of $r_0$ upon $d_{cl}$; this is plotted in Figure 5. The claimed observational relationship is that $r_0 \simeq 0.4\,d_{cl}$ and is plotted as the dashed straight line. None of the models produce a richness–clustering relationship as strong as the claimed dependence; only the OCDM model produce the large correlation length required by $R = 2$ bell clusters ($d_{cl} = 84\,h^{-1}$ Mpc), but it fails to give the observed amount of clustering for APM–like clusters. However, it is wise to sound a note of caution concerning the reliability of observational results for $R = 2$ Abell clusters: it is not clear whether available cluster samples are large enough to provide meaningful statistical estimates of the distribution of such rare objects.

We can also use our simulations to check the validity of simple theoretical calculations of the cluster correlations based on the assumption that they form at peaks in the primordial density field. According to the Kaiser (1984) model, clusters thus defined provide a biased tracer of the linear mass correlations so that

$$\xi_{cc}(r) \simeq b_{cl}^2 \xi(r), \qquad (6)$$

where $\xi(r)$ is the autocorrelation function of the density field (filtered on an appropriate scale). In the Kaiser calculation, $b_{cl}$ is constant on large scales and for very high peaks. A constant multiplier seems to be roughly what is required to explain the observed correlations if galaxies trace the mass distribution. Using (6) as a definition of $b_{cl}$, we can calculate the dependence of $b_{cl}$ upon scale $r$ for our different models. The results are shown in Figure 6. Note that the level of bias is very different in the different models and $b_{cl}$ is not constant in $r$ for any of them, except within the filtering scale and, only for OCDM, for the richest clusters. These results provide a warning to those who insist on using a constant bias factor to relate mass correlations to those of rich clusters. The fact that $b_{cl}$ varies with scale is not surprising: it is a generic feature of models with a *local* bias (e.g. Coles 1993). What is perhaps surprising is that $b_{cl}$ is an increasing function of scale beyond 20 $h^{-1}$ Mpc when Coles (1993) shows that it should be a non–increasing function as long as the fluctuations are Gaussian and the mass correlations are still linear. These conditions are violated by the significant dynamical evolution of the clusters away from their initial positions.

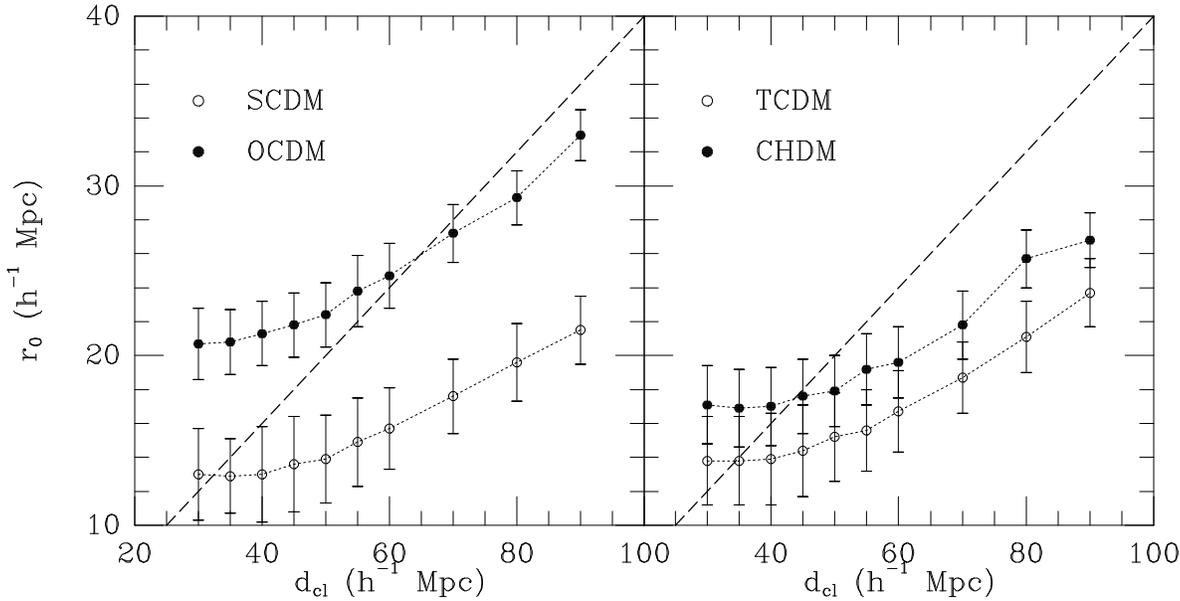

**Figure 5.** The $d_{cl}$–$r_0$ relation for the cluster simulations. The plotted errorbars correspond to one standard deviation in the unweighted least–square fit to $\xi_{cc}(r)$ in the scale–range 10–40 $h^{-1}$ Mpc. The dashed line is the relation $r_0 = 0.4\,d_{cl}$.

## 4 CONCLUSIONS

The first conclusion of this work is that our method based on the Zel'dovich approximation is indeed a fast and accurate way to generate simulated cluster samples. We have generated an ensemble of 20 simulations in a 640 $h^{-1}$ Mpc box for each of four cosmological scenarios. Our results agree with less complete analyses based on "exact" $N$–body simulations, which provide a less accurate sampling of long wavelength modes.

We confirm that the CDM model with $b = 1$ does not produce clusters with a large enough correlation length to match the observations. Our open CDM model seems to have too large a clustering length. CDM models with slight tilt ($n = 0.8$) shows clustering properties marginally consistent with observations; a larger tilt ($n = 0.7$) gives the right amount of clustering. The mixed CHDM model seems to be in general agreement with the observations.

Our conclusion for the open model contrasts with that of Bahcall & Cen (1992): they claim that an open CDM model provides a good fit to the observations. In fact, Bahcall & Cen (1992) use a model with a different value of $h$ from ours. One would expect the clustering length to scale roughly with $(\Omega_0 h)^{-1}$ so that our model with $h = 1$ should give a clustering length approximately 2.5 times that in the standard CDM model. This is, in fact, what we find. Bahcall & Cen (1992), however, use $h = 0.5$ which should give a clustering length about five times the standard CDM results, in clear conflict with the observations. Instead they find only about a factor two. The likely explanation for this is that their simulation (for they use only one) fails to sample adequately the long wavelength modes in this model. That is to say, their version of the OCDM model as too much large scale power to be represented accurately by their computational box.

We find the relationship between cluster richness and correlation length to be rather weak in all the models. This result agrees with those of Croft & Efstathiou (1993) and Mann et al. (1993), but disagrees with that of Bahcall & Cen (1992); the latter authors claim that the OCDM model produces a strong richness–clustering relationship. Croft & Efstathiou (1993) have made some suggestions as to the possible discrepancies between their simulations and those of Bahcall & Cen (1992); the force resolution is different in the two simulations, for example. We find it difficult to understand how such effects could influence correlations on the largest scales. One clue might be the discrepancy noted above for the clustering length in the open CDM model. In any case, all our models fail to reproduce the very high clustering amplitude of $R \geq 2$ Abell clusters reported by Postman et al. (1986). However, since we have only used a crude recipe for assigning masses to clusters, these conclusions are only indicative.

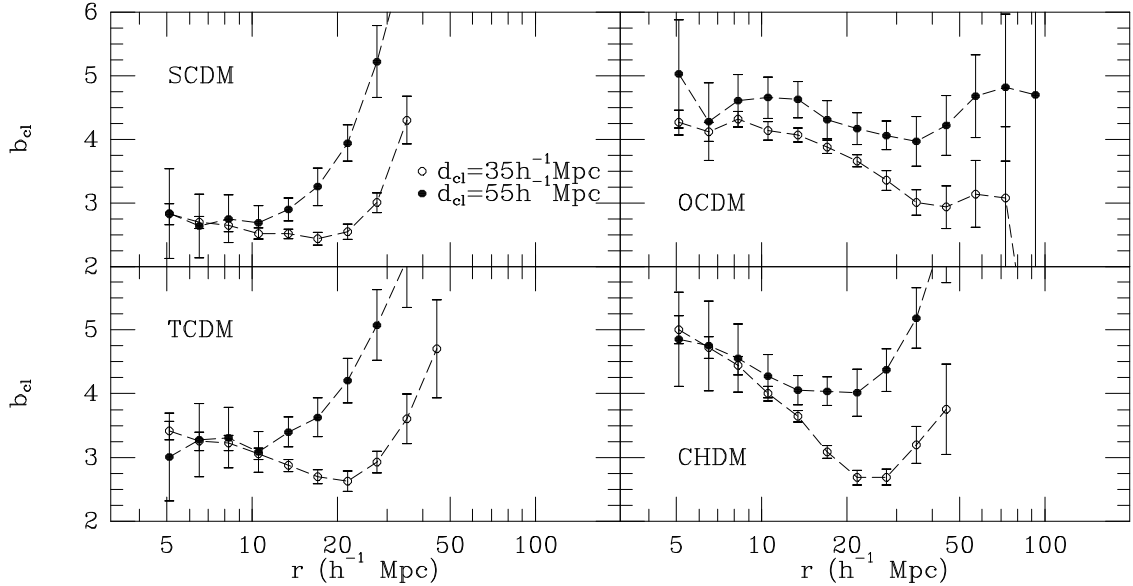

**Figure 6.** The cluster biasing factor plotted as a function of the scale for $d_{cl} = 35\,h^{-1}$ Mpc and $d_{cl} = 55\,h^{-1}$ Mpc cluster populations (open and filled dots, respectively). For each model, $b_{cl}$ is plotted up to the scale where the corresponding cluster correlation function first crosses zero.

KR93 have suggested the use of the zero–crossing of the cluster–cluster two–point correlation function as a test of models. Although there are differences in the $r_c$ values between different models, nevertheless we should remember that we tested the weak clustering regime in the ideal case of large samples with periodic boundary conditions and without any observational bias. Whether the $r_c$ statistics remains robust when including realistic sample geometries and selection effects remains to be checked.

We have also demonstrated that none of our models produces clusters which can be a locally biased subset of the underlying linear density field. The use of constant bias factors in cosmological contexts is widespread, particularly when the aim is to relate the velocities of galaxies or clusters to mass fluctuations and thus to estimate $\Omega_0$. Our results clearly suggest that one should tread very carefully when doing this, because there is no reason to expect the bias factor to be constant, even on the large scales probed by clusters. It also casts doubt on simple scaling of cluster power–spectra to provide the mass power–spectrum on large scales (e.g. Peacock & West 1992; Peacock & Dodds 1993). We shall look at the problems caused by the scale dependence of $b_{cl}$ in future work.

Indeed, in order to perform a rigorous test of cosmological models using cluster correlations one would need to reproduce exactly the same sampling and selection effects in the simulations as in the observational catalogues. In future work we shall be using our Zel'dovich catalogues to generate more realistic mock cluster samples so that we can assess the impact of such effects on $\xi_{cc}(r)$ as well as on other higher–order statistics. The existence of these possible biases is further motivation for our method, for only by careful estimation of the statistical uncertainty across a large ensemble can one perform a rigorous test of theory against observations.

**Acknowledgments**

We thank A. Klypin for supplying the KR93 simulation results. Peter Coles receives an SERC Advanced Fellowship; he is also grateful to the Dipartimento di Astronomia, Università di Padova for their hospitality during a visit when this paper was begun and to the Consiglio Nazionale delle Ricerche for financial support. LM acknowledges the Italian MURST for partial financial support, and the CINECA Computer Center (Bologna, Italy) for the use of computing facilities.